\documentstyle[12pt]{article}
\def\bold#1{\setbox0=\hbox{$#1$}%
      \kern-.025em\copy0\kern-\wd0
      \kern.05em\copy0\kern-\wd0
      \kern-.025em\raise.0433em\box0 }
\def\l{\left}
\def\r{\right}
\def\nsp{\noindent}
\def\nn{\nonumber}
\def\eea{\end{eqnarray}}
\def\bea{\begin{eqnarray}}
\def\eeas{\end{eqnarray*}}
\def\beas{\begin{eqnarray*}}
\def\ee{\end{equation}}
\def\be{\begin{equation}}
\def\bdm{\begin{displaymath}}
\def\edm{\end{displaymath}}
\def\fr{\frac}

\def\fks{{f_K^2}}
\def\sfrc{\fr{\sin^2 F}{r^2}}

\def\dag{^\dagger}
\def\tr{\mbox{Tr}}
\def\Tr{\mbox{Tr}}

\def\fpi2{\mbox{F$_\pi$}^2}

\def\mpi2{{m_\pi}^2}
\def\mk{m_K}
\def\mk2{{m_K}^2}

\def\fk2{\mbox{F$_K$}^2}

\def\fpis{{f_\pi^2}}

\def\skp{\epsilon}
\def\mss{\mu_{s,0}}
\def\msk{\mu_{s,1}}
\def\mvs{\mu_{v,0}}
\def\mvk{\mu_{v,1}}

\renewcommand{\thefootnote}{\fnsymbol{footnote}}
\begin{document}
\begin{titlepage}
\begin{center}
\hfill FNT/T-95/21

\vspace*{2.0cm}
{\large\bf HYPERON POLARIZABILITIES \\
 IN THE BOUND STATE SOLITON MODEL}
\vskip 1.5cm

{Carlo GOBBI$^a$,
Carlos L. SCHAT$^b$
and Norberto N. SCOCCOLA$^{b,c}$\footnote[2]{Fellow
of the CONICET, Argentina.} }
\vskip .2cm
{\it
$^a$ Department of Theoretical and
Nuclear Physics, University of Pavia,\\
and INFN, Sezione di Pavia, via Bassi 6, I-27100 Pavia, Italy.\\
$^b$ Physics Department, Comisi\'on Nacional de Energ\'{\i}a At\'omica,
          Av.Libertador 8250, (1429) Buenos Aires, Argentina. \\
$^c$ INFN, Sezione di Milano, via Celoria 16, I-20133 Milano, Italy.\\  }

\vskip 2.cm
August 1995
\vskip 2.cm
{\bf ABSTRACT}\\
\begin{quotation}
A detailed calculation of electric and magnetic static polarizabilities
of octet hyperons is presented in the framework of the bound state soliton
model. Both seagull and dispersive contributions are considered,
and the results are compared with different model predictions.
\end{quotation}
\end{center}
\end{titlepage}

\renewcommand{\thefootnote}{\arabic{footnote}}

\section{Introduction}

The electromagnetic polarizabilities are quantities of fundamental
interest in the understanding of hadron structure \cite{Pet81}.
They characterize the dynamical response of the hadron to
external electromagnetic fields. While a rather
large amount of work has been devoted, both theoretically and
experimentally, to the study of the proton and neutron polarizabilities
(see e.g. Refs.\cite{Mac95,Lvo93} for recent experimental and
theoretical reviews, respectively) very little is known about the
hyperon polarizabilities. However, with the advent of
hyperon beams at FNAL and CERN, the experimental situation
is likely to change. In particular, $\Sigma$ hyperon polarizabilities
will be soon measured in the Fermilab E781 SELEX experiment
\cite{FNAL94,Moi94}. This has triggered a number
of theoretical investigations in different hadron models.
In fact, predictions have been made in the framework
of the non-relativistic quark model (NRQM) \cite{LM92} and
heavy baryon chiral perturbation theory (HBCPT) \cite{BKKM92}.
As it is well-known the above-mentioned models have a few problems
in describing baryon magnetic polarizabilities.
Within the NRQM the large
diamagnetic contribution to the nucleon magnetic
polarizability  is rather difficult to understand.
In the case of HBCPT predictions are not expected
to be very accurate unless the
contributions due to P-wave excitations ($\Delta$-like),
which are of higher order in the chiral expansion, are included.

It is, therefore, interesting to attempt a description based on a completely
different point of view, like the one given by the topological (Skyrme)
soliton model.
Within the chiral soliton model only electric hyperon
polarizabilities have been so far studied \cite{Sch93}.
In the present work we will explore the static electric and
magnetic polarizabilities using the bound-state soliton
model\cite{CK85,SNNR88}, which has already given good results for
hyperon magnetic moments and mean square radii\cite{OMRS91,GBR}.

This article is organized as follows: In Sec.2 we introduce
the soliton model effective action in the presence of e.m. fields. In Sec.3
we briefly discuss how hyperons are described in the bound
state approach and in Sec.4 we calculate the static electric
and magnetic polarizabilities. Numerical results are reported
in Sec.5, while Sec.6 contains the conclusions. In Appendix A we estimate the
dispersive contributions to the hyperon electric polarizability. In Appendix B
and in Appendix C we give the explicit expressions of the (elementary)
polarizabilities and magnetic moments respectively.

\section{The effective action in the presence of electromagnetic fields}

Our starting point is a gauged effective chiral action with an
appropriate symmetry breaking term. It has the form
\be
\Gamma = \Gamma_{SK} + \Gamma_{an} + \Gamma_{sb}
\label{lag}
\ee
where $\Gamma_{SK}$ is the gauged Skyrme action
\be
\Gamma_{SK} =
\int d^4 x \Big\{ {f^2_\pi \over 4} \Tr\left[ D_\mu U (D^\mu U)^\dagger \right]
+
 {1\over{32 \skp^2}}
 \Tr\left[ [U^\dagger D_\mu U , U^\dagger D_\nu U]^2\right] \Big\} \, .
\ee
Here $f_\pi$ is the pion decay constant ( $=93 \ MeV$ empirically),
$\epsilon$ is a
dimensionless constant (the so-called Skyrme parameter) and $U$ is the  $SU(3)$
valued chiral field. The covariant derivative is defined as
\be
D_\mu U = \partial_\mu U + ie\ A_\mu \ [Q, U ]
\ee
where $A_\mu$ is the electromagnetic field and $Q$
the electric charge operator
\be
Q = {1\over2} \ \left[ \lambda_3 + {1\over{\sqrt{3}}} \lambda_8
\right] \, .
\ee
Moreover, $e$ represents the elementary electric charge. Since we are using
Gaussian units throughout the paper, in the following we adopt $e^2=1/137$.

$\Gamma_{an}$ is the gauged Wess-Zumino action which
for the electromagnetic case we are interested in reads\cite{Wit83}:
\bea
\Gamma_{an} &=& - {i N_c \over{240 \pi^2}}
\int \Tr[ (U^\dagger d U)^5 ] \nn \\
& & - {N_c\over{48\pi^2}}
\int d^4 x \ \epsilon^{\mu\nu\rho\sigma} \left.\Big\{
e A_\mu \ \Tr \left[ Q \left( L_\nu L_\rho L_\sigma -
                           R_\nu R_\rho R_\sigma \right) \right]
\right.\nn \\
& & \left. \qquad - i e^2 A_{\mu} \partial_\nu A_\rho \
\Tr \left[2\ Q^2 (L_\sigma - R_\sigma) +
    Q U^\dagger Q U L_\sigma
   -  Q U Q U^\dagger R_\sigma \right] \right\} \label{an} \, ,
\eea
where $L_\mu = U^\dagger \partial_\mu U$,
$R_\mu = U \partial_\mu U^\dagger$ and $N_c$ is the number of colors.
Finally, $\Gamma_{\rm sb}$ is the symmetry breaking term~\cite{RS91}:
\bea
\Gamma_{sb} & = &\int d^4x \left\{
 { f_\pi^2 m_\pi^2 + 2 f_K^2 m_K^2 \over{12} }
 \Tr \left[ U + U^\dagger - 2 \right] \right.
\nonumber \\
& & \qquad \left.
+ \sqrt{3}  { f_\pi^2 m_\pi^2 - f_K^2 m_K^2 \over{6} }
\Tr \left[ \lambda_8 \left( U + U^\dagger \right) \right] \right.
\nonumber \\
& & \qquad
 \left.
+ { f_K^2 - f_\pi^2\over{12} }
\Tr \left[ (1- \sqrt{3} \lambda_8)
\left(
U (D_\mu U)^\dagger D^\mu U + U^\dagger D_\mu U (D^\mu U)^\dagger \right)
\right] \right\} \, ,
\label{sb}
\eea
\noindent
where $f_K$ is the kaon decay constant and
$m_\pi$ and $m_K$ are the pion and kaon masses respectively.

For our purposes, the effective action can be more conveniently
written as
\be \label{gafin}
\Gamma = \Gamma^{strong} +
        \Gamma^{lin} + \Gamma^{quad} \, ,
\ee
where we have singled out the contributions linear and quadratic in the
e.m. field:
\bea
\Gamma^{lin} & = & \int d^4x \ e \ A_\mu J^\mu \, , \\
\Gamma^{quad} & = & -\int d^4x \ e^2 \ A_\mu \ G^{\mu\nu} \ A_\nu \, .
\eea
Here:
\bea
J^\mu &=& i {f_\pi^2\over2} \Tr\left\{ Q (L^\mu + R^\mu) \right\} \nn \\
      & &   + i {f_K^2 - f_\pi^2\over{12}}
         \Tr\left\{(1-\sqrt{3} \lambda_8)\,
([U,Q]L^\mu - L^\mu [U^\dagger,Q]+[U^\dagger,Q]R^\mu -R^\mu[U,Q])
 \right\} \nn\\
    & & - {i\over{8\skp^2}} \Tr\left\{ Q \left( [L_\nu,[L^\mu,L^\nu]]
                         + [R_\nu,[R^\mu,R^\nu]] \right) \right\} \nn\\
    & & - {N_c\over{48\pi^2}} \epsilon^{\mu\nu\rho\sigma}
              \Tr\left\{ Q \left( L_\nu L_\rho L_\sigma -
                            R_\nu R_\rho R_\sigma \right) \right\} \, ,
\eea
and
\bea
G^{\mu\nu} &=& - g^{\mu\nu}
         \left[ {f_\pi^2\over4} \Tr P^2 +
         {f_K^2 - f_\pi^2\over{12}} \Tr\left\{ (1-\sqrt{3} \lambda_8)
         (P^2 U^\dagger + U P^2) \right\} \right] \nn \\
  & & + {1\over{8\skp^2}}
      \Big[ g^{\mu\nu} h_\alpha^\alpha - h^{\mu\nu} \Big] \nn \\
  & & + {i N_c\over{48\pi^2}} \epsilon^{\mu\nu\rho\sigma}
        \Tr\left[ (2Q^2+QU^\dagger Q U) L_\sigma -
                 (2Q^2+Q U Q U^\dagger) R_\sigma \right] \partial_\rho \, ,
\eea
\nsp
where the following definitions have been used:
\bea
P &=& Q - U^\dagger \ Q \ U  \label{P} \, , \\
h_{\mu\nu} &=& \Tr\left[ P L_\mu P L_\nu - P^2 L_\nu L_\mu \right] \label{h}\,
{}.
\eea

In Eq.(\ref{gafin}) $\Gamma^{strong}$ is the action in the absence
of the electromagnetic field. It describes the strong interactions
that give rise to the hyperon. In the next section it will be
treated following the usual steps of the bound state model.

\section{Hyperons in the bound state soliton model}

The bound state soliton model has been discussed in great detail
in the literature (see e.g. Refs.\cite{CK85,SNNR88}). Therefore,
in this section we will only present the main features
of the model. Following Ref.\cite{CK85} we introduce the Callan--Klebanov
ansatz
\be   \label{ansatz}
U=\sqrt{U_\pi} \, U_{K} \,  \sqrt{U_\pi} \ ,
\ee
where
\bea
U_K \ = \ \exp \left[ i\fr{\sqrt2}{{f_K}} \left( \begin{array}{cc}
                                                        0 & K \\
                                                        K\dag & 0
                                                   \end{array}
                                           \right) \right] \ , \
                                           \ \ \
K \ = \ \left( \begin{array}{c}
                   K^+ \\
                   K^0
                \end{array}
                           \right),
\eea
and $U_\pi$ is the soliton background field written as
a direct extension to $SU(3)$ of the $SU(2)$ field $u_\pi$, i.e.,
\bea
U_\pi \ = \ \left ( \begin{array}{cc}
                       u_\pi & 0 \\
                       0 & 1
                    \end{array}
                               \right ) \ ,
\eea
with $u_\pi$ being the conventional hedgehog solution
$ u_\pi=\exp[i \bold \tau \cdot \hat{\bold r}F (r) ]$.

According to the usual procedure, one replaces the ansatz
(\ref{ansatz}) in the effective action $\Gamma^{strong}$ and
expands up to the second order in the kaon field. The
resulting Lagrangian density can then be written as the sum of a pure
$SU(2)$ Lagrangian depending on the chiral field only and an effective
Lagrangian, describing the interaction between the soliton and the kaon
fields. The minimization of the first term determines the static soliton
profile $F(r)$ (chiral angle). The minimization of the second one leads
to an eigenvalue equation for the time-dependent meson field $K$ in the
static potential field of the $SU(2)$ soliton.
The bound state solutions to this wave equation represent stable hyperon
states. Due to the spin--isospin structure of the soliton this
eigenvalue equation becomes separable if a mode decomposition of the kaon
field in terms of the
grand spin $\bold \Lambda=\bold L + \bold T$ (where $\bold L$ represents the
angular momentum operator and $\bold T$ is the isospin operator) is
performed.
As shown in Refs.\cite{CK85,SNNR88} the lowest bound state is found in the
$(\Lambda,l)=(1/2,1)$
channel. The different octet and decuplet baryons are obtained by putting
the corresponding number of kaons in this bound state. However,
by naively adding $|{\cal S}|$ times the bound state energy $\omega$ to
the soliton mass one obtains only the centroid mass of the hyperons
with strangeness ${\cal S}$.
The splittings among hyperons with different spin and/or isospin are given
by the rotational corrections, introduced according to the time--dependent
rotations:
\bea
u_\pi & \to & A u_\pi A^\dagger\, ,    \nonumber \\
K     & \to & A K \, ,
\eea
with $A=A(t)$ being the SU(2) rotation matrix.
This transformation adds an extra term to the Lagrangian
which is of order $1/N_c$. Therefore, within our approximations
the strong hamiltonian reads
\be
H^{strong} = M_{sol} + |{\cal S}| \omega + {1\over{2\Theta}}
           (\vec J^c + c \vec J^K)^2 \, .
\ee
Here, $\Theta$ is the soliton moment of inertia,
and $c$ is the hyperfine splitting constant (its
explicit form for the cases of interest in this paper can be easily
obtained from the general form given  in Ref.~\cite{RRS92}).
Moreover, $\vec J^c$ and $\vec J^K$ are the collective and
bound kaons angular momentum operators, respectively.
Taking matrix elements of these operators between the different
octet and decuplet hyperon states we obtain their corresponding
masses in the absence of e.m. fields
\be
M_{I,J,{\cal S}} = M_{sol} + \omega \vert {\cal S} \vert +
{1\over{2\Theta}} \Big[ c J(J+1) + (1-c) I(I+1) +
{c(c-1)\over4}  \vert {\cal S} \vert (\vert {\cal S} \vert + 2 ) \Big] \, .
\ee
Here, $I$, $J$ and ${\cal S}$ are the isospin, spin and strangeness
hyperon quantum numbers respectively.

\section{The hyperon static polarizabilities}

In this paper we will be concerned with the static polarizabilities only,
defined through the shift in the particle's energy due to the presence of
an external constant electric and magnetic field as:
\bea
\delta M &=& - \fr{1}{2} \ \alpha \ E^2
- \fr{1}{2} \ \beta \ B^2 \label{shift}  \, .
\eea
The electric $\alpha$ and magnetic $\beta$ polarizabilities
characterize the dynamical response to the external electromagnetic fields.
In the following we will study the shifts in $E^{2}$ and $B^{2}$ separately,
by a proper choice for the electromagnetic potential $A^{\mu}$.
As it is clear from the form of the interaction  (\ref{gafin}),
there will be in principle two contributions to the static polarizabilities,
one coming from the term quadratic in $A^\mu$,
known as the {\it seagull contribution}
and one coming from second order perturbation theory applied to
the term linear in $A^\mu$, the so called {\it dispersive contribution}.

\subsection{The static electric polarizability}

The energy shift proportional to  $E^2$ can easily be obtained from
(\ref{gafin}) by adopting
a potential $A^\mu$ of the form
\be
A^\mu = (A_0, 0) \, , \qquad  \qquad A_0=-zE
\ee
which corresponds to a constant electric field $E$
along the $z$-axis. Using the definitions (\ref{P},\ref{h}),
the seagull contribution can be expressed as
{\small
\bea  \label{epol}
\alpha_s & =&   \fr{e^2}{2} \int d^3r \Bigg\{ z^2 \left[ \fpis \tr(P^2)
+ \fr{1}{2 \skp^2} h_{ii} +  {f_K^2 - f_\pi^2\over{3}} \Tr\left\{ (1-\sqrt{3}
\lambda_8)
         (P^2 U^\dagger + U P^2) \right\}
 \right] \Bigg\}  \, . \nn \\ & &
\eea
}
It should be noticed that in deriving Eq.(\ref{epol}) we have
assumed that the seagull contributions to the Hamiltonian are
simply equal to the seagull contributions to the Lagrangian, with the
opposite sign. There has been recently
some controversy about this point. In Refs.\cite{Lvo93,SU94}
it has been argued that on general grounds in a field theory
electric seagull contributions to the hamiltonian should vanish.
However, as discussed in Ref.\cite{SC95} this is not
the case when the degrees of freedom are restricted to be
in some collective subspace. In fact, in that reference
it has been explicitly shown that the procedure above is
completely valid when the Skyrme model in the presence of
a constant electric field is treated by introducing
collective coordinates.

The dispersive contribution $\alpha_d$
is determined by matrix elements describing transitions
between the particular octet state under investigation and negative parity
excited states. In general, $\alpha_d$ is believed to be much smaller
than $\alpha_s$\cite{Che87}. For this reason
we will not consider it further in our discussion.
An estimate of the approximation
introduced in this way is discussed in Appendix A.

Finally, we note that $\alpha_s$ contains no contributions coming
from the anomaly term (\ref{an}), because of the antisymmetric tensor
$\epsilon^{\mu\nu\rho\sigma}$.

Introducing the adiabatically rotated bound state ansatz in $\alpha_s$,
one obtains the following operator form
\bea \label{ae}
\alpha_s &=& \left[ \gamma_1^{(e)} + \gamma_2^{(e)}
\left(R_{33}\right)^2  \right.
+ \gamma_3^{(e)} |{\cal S}|   \nn \\
&  & \ \ \ \ \ \ \ \ \ \ + \left. \gamma_4^{(e)} |{\cal S}|
\left(R_{33}\right)^2
+ \gamma_5^{(e)}  \ J^K_a R_{3a}
+ \gamma_6^{(e)} \ J^K_3 R_{33}
\right]
\eea
where we have absorbed $e^2$ in  the elementary polarizabilities
$\gamma_i^{(e)}$ ($i=1, \dots, 6$), which depend
on the radial part of the soliton and bound kaon wavefunctions only. Their
explicit expressions are listed in Appendix B.
$R_{ab}$ are the rotation matrices defined by
\be
R_{ab} = {1\over2} \Tr \left[ \tau_a \ A \ \tau_b \ A^\dagger \right] \, .
\ee

In order to get the expressions for hyperon polarizabilities we have
to evaluate the matrix elements of the operators
appearing in (\ref{ae}) between
hyperon states. This is done using standard angular momentum techniques.
For the ground state octet baryons we obtain
\bea
\alpha_s(\Lambda) &=& \gamma_1^{(e)} + \gamma_3^{(e)} + \fr{1}{3}
\l(\gamma_2^{(e)} + \gamma_4^{(e)}\right) \label{25}\, ,  \\
\alpha_s(\Sigma_0) &=& \gamma_1^{(e)} + \gamma_3^{(e)} + \fr{1}{3}
\l(\gamma_2^{(e)} + \gamma_4^{(e)}\right)\, ,  \\
\alpha_s(\Sigma_\pm) &=& \gamma_1^{(e)} + \gamma_3^{(e)} + \fr{1}{3}
\l(\gamma_2^{(e)} + \gamma_4^{(e)}\right)
\pm \fr{1}{2} \l( \gamma_5^{(e)} + \fr{1}{3} \gamma_6^{(e)} \right) \, , \\
\alpha_s(\Xi_{\mbox{{\footnotesize{\underline 0}}}}) &=& \gamma_1^{(e)}
+ 2 \gamma_3^{(e)} + \fr{1}{3}\l(\gamma_2^{(e)} + 2 \gamma_4^{(e)}\r)
\pm \fr{2}{3}\l( \gamma_5^{(e)} +  \fr{1}{3} \gamma_6^{(e)} \right) \label{28}
\, .
\eea

\subsection{The static magnetic polarizability}

We proceed along similar lines to derive the static magnetic polarizability.
In this case we adopt the vector potential
\be
A^\mu = (0, -{1 \over 2} \bold r \times \bold B)
\ee
appropriate for a constant magnetic field $B$ along the $z$-axis,
$\bold B = B \hat z$.
Now we have to take into account both seagull and dispersive
contributions. In fact, the Hamiltonian form of (\ref{gafin}) reads
\be \label{H}
H = H^{strong} +
        H^{lin} + H^{quad} \, ,
\ee
where, again, all the contributions from $L^{an}$ vanish for symmetry
reasons. \\
The quadratic part yields the seagull contribution as for the electric
polarizability. Its explicit form, in terms of the operators $P$ and $h_{\mu
\nu}$ is
\bea
\label{betas}
 \beta_s \! \! &  \! \! = \! \! & \! \!
-\fr{e^2}{8} \int d^3r \Bigg\{ (r^2 - z^2)
\left[ f_\pi^2 \tr(P^2) + {f_K^2 - f_\pi^2\over{3}}
\Tr\left\{ (1-\sqrt{3} \lambda_8)
         (P^2 U^\dagger + U P^2) \right\} \right.
\Bigg] + \nn \\
& &  \ \ \ \ \ \ \ \ \ \ \ \qquad
+ {1\over{2\skp^2}} \left[ r_i r_j h_{ij} + r^2 h_{33}
- r_i r_3 (h_{i3} + h_{3i})- (r^2 - z^2) h_{00} \right] \ \Bigg\}\, .
\eea
A lengthy calculation shows that $\beta_s$ has the same operatorial
form of $\alpha_s$:
\bea
\beta_s &=& \l[ \gamma_1^{(m)} + \gamma_2^{(m)}
\left(R_{33}\right)^2
+ \gamma_3^{(m)} |{\cal S}| \right.   \nn \\
& & \ \ \ \ \ \ \ \ \ \ \left. + \gamma_4^{(m)} |{\cal S}|\l(R_{33}\right)^2
+ \gamma_5^{(m)}  \ J^K_a R_{3a}+ \gamma_6^{(m)}
\ J^K_3 R_{33} \right]  \, ,
\label{betadi}
\eea
where the operators involved have to be evaluated again between  the same
hyperon states. Therefore for the seagull part of the magnetic polarizability
we obtain again formal expressions similar to Eqs.(\ref{25})--(\ref{28}),
with the elementary polarizabilities $\gamma_i^{(e)}$ replaced by
$\gamma_i^{(m)}$ ($i=1,\ldots,6$).
Their explicit form is reported in Appendix B.

We must stress that in deriving Eq.(\ref{betas}) we have again
implicitly assumed
$H^{quad}=-L^{quad}$. This is correct up to a small contribution coming from
$L^{lin}$, proportional to the ratio $(\mu_s/M_N)^2$, where $\mu_s$ and
$M_N$ are the isoscalar magnetic moment operator and nucleon mass,
respectively.
For a more complete discussion, see for instance Ref.\cite{SW89}.

The dispersive contribution arises from the term $H^{lin}$ in (\ref{H}).
Using second order perturbation theory we get

\be
\beta_d^H =  {e^2\over{2 M_N^2}} \sum_{H'\neq H} {{\vert \langle H \vert
\mu_3 \vert H' \rangle \vert^2   }\over {m_{H'} - m_{H}} }
\label{betad}
\ee
where the indices $H$ and $H'$ refer to different hyperon states.
In writing Eq.(\ref{betad}) we have used the explicit form
of $H^{lin}$ for the particular case of a constant magnetic
field $B$ along the z-axis
\be
H^{lin} = - {e\over{2 M_N}}\ B \ \mu_3
\ee
where $\mu_3$ is the magnetic moment operator.
It can be written as a sum of an isoscalar and isovector
part as follows\cite{KM90,OMRS91}
\bea
\mu^3  &  =  &  \mu^3_s +\mu^3_v \, , \\
\mu^3_s &  = &  \mu_{s,0} J^c_3 + \mu_{s,1} J^K_3 \, ,  \\
\mu^3_v  &  = & -2\, \l( \mu_{v,0}  +\mu_{v,1} \vert {\cal S} \vert \right)
                 R_{33}\, .
\eea
The explicit expressions of the elementary magnetic moment operators
$\mu_{s,i}$ and $\mu_{v,i}$ are reported in Appendix C.
In terms of them the relevant matrix elements are
\bea
< \Lambda | \mu_3 | \Sigma_0 > &=& - {2\over3}
\left[ \mvs + \mvk \right] \, , \\
< \Lambda | \mu_3 | \Sigma^*_0 > &=& {2\sqrt{2}\over3}
\left[ \mvs + \mvk \right] \, , \\
< \Sigma_0 | \mu_3 | \Sigma^*_0  > &=& {\sqrt{2}\over3}
\left[ \mss - \msk \right] \, ,  \\
< \Sigma_\pm | \mu_3 | \Sigma^*_\pm  > &=&
{\sqrt{2}\over3} \left[ \mss - \msk \pm
(\mvs + \mvk)  \right] \, , \\
< \Xi_{\mbox{{\footnotesize{\underline 0}}} }
 | \mu_3 | \Xi^*_{\mbox{{\footnotesize{\underline 0}}} } > &=&
{\sqrt{2}\over3} \left[ \mss - \msk \pm {4\over3}
(\mvs + 2 \mvk)  \right]  \, .
\eea
Note that for each octet hyperon only a few matrix elements are non-vanishing.

\section{Numerical results and discussion}

In order to estimate the uncertainties intrinsic to our approach we
have performed numerical calculations adopting two different sets of
parameters, namely
\bea
{\mbox {\rm SET  I}}:
& & f_\pi = 93 MeV , \ \epsilon = 4.26 \ ,
                                                        \nn \\
{\mbox {\rm SET  II}}:
& & f_\pi = 54 MeV , \ \epsilon = 4.84 \ .
                                                         \nn
\eea
In both cases we use the empirical values
$m_\pi=138 \ MeV$, $m_K=495\ MeV$ and $f_K/f_\pi=1.22$.
In the first set of parameters we
have taken the empirical value of $f_\pi$. In
the second set we have taken the value of $f_\pi$ that
fits the empirical nucleon mass. In both sets $\epsilon$ is
adjusted to reproduce the empirical $\Delta-N$ mass difference.

The results obtained with the sets of parameters given above
are summarized in Tables I-V.
In Tables I and II, we list the elementary
polarizabilities $\gamma_i$ for the electric and magnetic
cases, respectively. In Table III we give the elementary
magnetic moments needed to calculate the dispersive
contributions to the magnetic polarizabilities. These
values have already been given in Refs.\cite{OMRS91,PS92}
and are included here for the sake of completeness.
Finally, in Table IV and V we report our results for
the electric and magnetic static hyperon polarizabilities, respectively.
In the case of the magnetic polarizabilities we also
list the dispersive and seagull contributions separately.

Let us first discuss the values of the elementary
polarizabilities $\gamma_i$. We see that for both sets of
parameters the purely solitonic contributions $\gamma_1$ and
$\gamma_2$ are much larger than the others. This holds for both
the electric and magnetic cases. As a result of this behaviour,
we expect a rather small splitting between the seagull contributions
to the polarizabilities of the different baryons. This can be in
fact observed in Tables IV and V. We also note that the
values of $\gamma_i$ are rather strongly dependent on the
values of the input parameters used. At least, in the case of
$\gamma_1$ and $\gamma_2$ (which are, as already mentioned, the dominant
terms) this is to be expected. As well-known within the Skyrme model
these magnitudes are
basically proportional to the square of the nucleon isovector
radius\footnote{This relation holds strictly for the electric seagull term.
In the magnetic case there is a (numerically)
small correction.} which in turn is quite sensible to the choice
of parameters. SET I leads to the value
$<r_v^2> = 0.70\ fm^2$ while the value obtained with SET II
is $<r_v^2> = 1.08 \ fm^2$ as compared with the
empirical value $<r_v^2>_{emp} = 0.81 \ fm^2$.
This dependence on the parameters reflects, of course, on the
values of all the electric and diamagnetic hyperon polarizabilities.
On the other hand, the dispersive contributions to the magnetic
polarizabilities are much more stable under change of parameters.
This comes as a result of the compensation between the parameter
dependence of the numerator and denominator in Eq.(\ref{betad}).

It is interesting to compare our predictions with those obtained
in other models. Our results indicate a rather large $\Sigma^+$
electric polarizability. This is in agreement with the quark model
prediction of Ref.\cite{LM92},
${{\bar \alpha}_{\Sigma^+}}^{NRQM} = 20.8 \times 10^{-4} fm^3 $.
However, such a model predicts a rather small value for the
case of the $\Sigma^-$, namely
${{\bar \alpha}_{\Sigma^-}}^{NRQM} = 5.7 \times 10^{-4} fm^3 $.
Although we also predict a smaller value for the $\Sigma^-$ as
compared with that of the $\Sigma^+$, in our case the
splitting between both values is much smaller. As mentioned
above this is a direct consequence of the fact that in our
model the electric polarizabilities are completely dominated
by the purely solitonic contributions. Small splittings
have been also found in the Skyrme model within the framework
of the perturbative approximation to the $SU(3)$ collective
coordinate approach \cite{Sch93}\footnote{In Ref.\cite{Sch93} it has
been incorrectly stated that
one of the non-minimal photon coupling terms does not
contribute to the electric polarizability. As a matter of fact it does,
and almost completely compensates the contribution from the other
non-minimal term \cite{Sch95}. As a consequence of this cancellation the
results of Ref.\cite{Sch93} have to be modified.
In fact, the addition of the missing contributions
amounts to a roughly $\ 20 \%$ increase of all the polarizabilities
corresponding to the perturbative calculation (PT)
and a $\  40 \%$ increase of those corresponding to the
slow rotator approach (SRA). In both cases
the ratios taken with respect to the proton polarizability
remain almost unchanged.}.
The use of an exact diagonalization procedure\cite{YA88}
does not change the overall behaviour \cite{Sch95}. Only by the
introduction of a feedback from the collective $SU(3)$ rotation
on the soliton, that is using the so-called slow rotator approach (SRA),
 large splittings between the electric polarizabilities
of the different hyperons could be obtained. In such case, the
electric polarizabilities decrease with increasing (absolute)
values of strangeness. This behaviour is similar to the one obtained
in chiral perturbation theory\cite{BKKM92}. It should be noticed,
however, that in the SRA calculation of Ref.\cite{Sch93} this is
obtained at expenses of a rather small isovector radius
$<r^2_v> = 0.49 \ fm^2$.

To complete our discussion,
it is important to mention that for the case of the
nucleon, where the empirical value of the electric polarizability
$(\alpha_N)_{emp} = 12 \times 10^{-4} fm^3 $ is rather well established
(see Ref.\cite{Mac95} for a very recent determination and update
of the experimental situation), the Skyrme model predicts a somewhat
large value for SET I and too large for SET II. As in the case of the
soliton mass, however, there are indications that this might be cured
by next-to-leading order corrections\cite{Wal95}

We turn now to the magnetic polarizabilities. Due basically to the
dispersive contributions our results indicate rather large splittings
between the values corresponding to the different hyperons. We also
observe that since seagull contributions are overestimated for Set II
we obtain all negative values in that case. For Set I our predicted
$\Sigma^+$ magnetic polarizability agrees well with the one obtained
in the non-relativistic quark model\cite{LM92}. On the other hand, in
the case of the $\Sigma^-$ although we also obtain a diamagnetic
behaviour, our value is larger (in absolute value). The results
obtained by using baryon chiral perturbation are rather different
from ours. In should be noticed that such a calculation does not
include $P$-wave excitations ($\Delta$-like) since they are of higher
order in the chiral expansion. Therefore, predictions are not
expected to be as accurate as in the electric case.

\section{Conclusions}

In this paper we have presented a complete description of static
electric and magnetic polarizabilities of octet hyperons in the framework
of the bound state soliton model. In the electric case the seagull contribution
is dominant, while in the magnetic case both seagull and dispersive
contributions are relevant.

As shown by numerical calculation, the seagull contributions are always
dominated by the purely solitonic terms, $\gamma_1$ and $\gamma_2$.
These pieces determine the general pattern for electric polarizabilities,
where we obtain small splittings within the same set of parameters.
The structure  is richer in the magnetic polarizability case
because of the interplay between a large (negative) seagull part with
the relevant dispersive contribution.

Finally, we note that although some of our results
are in agreement with those of the non-relativistic quark model,
in general this is not the case. In addition, the calculations
performed in the framework of heavy baryon chiral perturbation theory
lead to still different predictions. In this situation, it
is clear that the future experimental data from FNAL and CERN could
be of great help to discriminate among the different existing models
of hyperons.

\vspace*{1.cm}

\section*{Acknowledgments}

We wish to thank S. Scherer and B. Schwesinger for useful
discussions and comments. NNS wishes to acknowledge discussions
with A. Rivas at the initial stage of this work.
Part of the
work reported here was done while two of them (CLS and NNS) were
participating at the INT-95-1 Session on ``Chiral Dynamics in Hadrons
and Nuclei" at the Institute for Nuclear Theory at the University of
Washington, USA. They wish to thank the organizers of the Session
for the invitation to participate at it and the Department of Energy, USA,
for partial financial support during that period.
CLS was partially supported by Fundaci\'on Antorchas and
CG by University of Pavia under
the Postdoctoral Fellowship Program.
\pagebreak
\section*{Appendix A}

In this Appendix we give an estimate of the value of the dispersive
contributions to the hyperon electric polarizability.
They are due to dipole transitions to the negative parity
excited states. As in the magnetic case, these contributions
are obtained using second order perturbation theory with the
linear terms $H^{lin}$ in the hamiltonian. In the case of
a static electric field the corresponding expression is
\be
\alpha_d^H = { 2 e^2 }\
\sum_{H'} {|< H' | d_3 | H >|^2 \over{m_{H'} - m_{H}} } \, ,
\ee
where $d_3$ is the third component of the dipole operator
\be
d_3 = \int dV \ z \ \rho^{em} \, .
\ee
We consider here the particular example
of the $\Lambda$ electric polarizability, where the
largest contribution is expected to be the one in which the
intermediate state is the $\Lambda(1405)$. Then, $H=\Lambda$ and
the sum over $H'$ is restricted to $H'=\Lambda(1405)$.
In this case we only need to consider the isoscalar
kaon contributions to $\rho^{em}$
\be
\rho^{em}(kaon) = {i\over2} f
\left[ K^\dagger \dot K - \dot K^\dagger K \right] - \lambda K^\dagger K \, ,
\ee
where
\bea
f &=& 1 + {1\over{4\skp^2 f_K^2}}
\left[ F'^2 + 2 {\sin^2 F\over{r^2}} \right] \, , \\
\lambda &=& - {N_c\over{8\pi^2 f_K^2}} {\sin^2 F\over{r^2}} \ F'\, .
\eea
Taking matrix elements of $\rho^{em}$ between $\Lambda (1405)$ and
$\Lambda$ we get
\be
< \Lambda (1405) | \rho^{em} | \Lambda > =
- \left[ f ({\tilde \omega} + \omega) + 2 \lambda \right] \
{{\tilde k}\ k\over{4\pi}} \  \hat r \cdot <\vec J>
\ee
\nsp
where $<\vec J>$
indicates the matrix elements
of the spin operator between the $\Lambda (1405)$ and $\Lambda$
spin states and (${\tilde \omega}$,
${\tilde k}$) and ($\omega$, $k$) are the kaon
eigenenergies and bound state radial wavefunctions
in the $(1/2,0)$ and $(1/2,1)$ channels, respectively.
Therefore
\be
< \Lambda (1405) | d_3 | \Lambda > = - \gamma
\ee
\nsp
where
\be
\gamma = {1\over6} \int dr \ r^3 \left[ f ({\tilde \omega} + \omega) +
2 \lambda \right] \ {\tilde k}\ k \, .
\ee
\nsp
To derive this expression the angular integral has been performed
and $<J_3>=1/2$ has been used. Replacing in the
expression for $\alpha$ we get that the contribution to the
$\Lambda$ electric polarizability due to dipole electric transition
to $\Lambda (1405)$ is
\be
\alpha^{\Lambda}_{d}
= { 2 \ e^2 \gamma^2 \over{m_{\Lambda (1405)} - m_{\Lambda}}}
\ee
Numerically, we find
\be
\alpha^{\Lambda}_{d}  = 0.54 \times 10^{-4} \ fm^3
\ee
for SET I and
\be
\alpha^{\Lambda}_{d}  = 1.08 \times 10^{-4} \ fm^3
\ee
\nsp
for SET II. As we see these values are much smaller than the seagull
contributions given in Table IV. Of course, it should be kept in mind
that this is just an estimation of the order of magnitude since a full
calculation should include all possible intermediate states.
Although here we have discussed only the case of the $\Lambda$ similar
results are expected for the other hyperons.

\section*{Appendix B}

The elementary electric polarizabilities are given by:
\bea
\gamma_1^{(e)} &=& \fr{16}{15} \pi\ e^2 \int dr \ r^4
\sin^2 F \l[ \fpis + \fr{1}{\skp^2} \l(F'^2 + \sfrc \right) \right] \, , \\
\nn \\
\gamma_2^{(e)} &=& - \fr{8}{15} \pi \ e^2\int dr \ r^4 \sin^2 F
\l[ \fpis + \fr{1}{\skp^2} \l(F'^2 + \sfrc \right) \right] \, ,    \\
\nn \\
\gamma_3^{(e)} &=& \fr{1}{15} \ e^2
\int dr \ r^4 \Bigg\{ k^2 (1 + 4 \cos^2 F) \nn \\
& & \ \ \ \ \ \ \ \ \ \ \ \ \ \ \ \ \ - \fr{1}{\skp^2 \fks}
\l[ \fr{9}{2} k^2 F'^2
\sin^2 F + 5 k^2 \fr{\sin^4 F}{r^2} -
\fr{5}{4} k^2 \l( F'^2 + 2 \sfrc \right) \right. \nn \\
& & \ \ \ \ \ \ \ \ \ \ \ \ \ \ \ \ \ \ \ \ \ \ \ \ \ \ \ \ \  - 2 k'^2
\sin^2 F - 2 \fr{k^2}{r^2} \sin^2 F \cos^2\fr{F}{2} (1+3 \cos F )  \nn \\
& & \ \ \ \ \ \ \ \ \ \ \ \ \ \ \ \ \ \ \ \ \ \ \ \ \ \ \ \ \ - 3 k k' F'
\sin 2F \Bigg] \ \Bigg\} \, ,    \\
\nn \\
\gamma_4^{(e)} &=& \fr{2}{15} \ e^2 \int dr \ r^4 \Bigg\{ k^2 \sin^2 F \nn \\
& & \ \ \ \ \ \ \ \ \ \ \ \ \ \ \ \ \ + \fr{1}{4 \skp^2 \fks} \l[ \fr{9}{2}
k^2 F'^2 \sin^2 F + 5 k^2 \fr{\sin^4 F}{r^2}  - 3 k k' F' \sin 2F  \right.
\nn \\
& & \ \ \ \ \ \ \ \ \ \ \ \ \ \ \ \ \ \ \ \ \ \ \ \  - 2 k'^2
\sin^2 F - 2 \fr{k^2}{r^2} \sin^2 F \cos^2\fr{F}{2} (1+3 \cos F ) \Bigg] \
\Bigg\}  \, ,  \\
\nn \\
\gamma_5^{(e)} &=& -\fr{2}{15} \ e^2
\int dr \ r^4 \Bigg\{ k^2 (1 - 4 \cos F) \nn \\
& & \ \ \ \ \ \ \ \ \ \ \ \ \ \ \ - \fr{1}{4 \skp^2 \fks} \l[ 16 k^2
\cos^2\fr{F}{2} \sfrc  - k^2 \l( F'^2 + 2 \sfrc \right) (1 - 4 \cos F)  \right.
\nn \\
& & \ \ \ \ \ \ \ \ \ \ \ \ \ \ \ \ \ \ \ \ \ \ \ \ \ \ \ \ \ + 24 k k' F'
\sin F \Bigg] \ \Bigg\} \,   ,   \\
\nn \\
\gamma_6^{(e)} &=& -\fr{8}{15}\ e^2 \int dr \ r^4
\Bigg\{ k^2 \cos^2 \fr{F}{2} \nn \\
& & \ \ \ \ \ \ \ \ \ \ \ \ \ \ \ \ \ + \fr{1}{4 \skp^2 \fks} \l[  k^2
\cos^2\fr{F}{2} \l( F'^2 + 4 \sfrc \right) + 3 k k' F' \sin F \right]
\ \Bigg\} \, .
\eea
\noindent
For the magnetic polarizability\footnote{Note that the expressions
of $\gamma_1^{(m)}$ and
$\gamma_2^{(m)}$ together with Eq.(\ref{betadi}) do not lead to
Eqs.(45-46) of Ref.\cite{SM92} which are in error. This
affects only the corresponding expressions for the $\Delta$
magnetic polarizabilities. The correct numerical values are,
however, very close to those quoted in such reference.}
we have:
\bea
\gamma_1^{(m)} &=& - \fr{2}{5} \pi \ e^2 \int dr \ r^4
\sin^2 F \l[ \fpis + \fr{1}{\skp^2} \l(F'^2 + \fr{1}{6}\sfrc \right)
\right] \, ,  \\
\nn \\
\gamma_2^{(m)} &=& - \fr{2}{15} \pi \ e^2 \int dr \ r^4 \sin^2 F
\l[ \fpis + \fr{1}{\skp^2} \l(F'^2 + \fr{7}{2}\sfrc \right) \right] \,  ,  \\
\nn \\
\gamma_3^{(m)} &=& -\fr{1}{30} \ e^2
\int dr \ r^4 \Bigg\{ k^2 (2 + 3 \cos^2 F) \nn \\
& & \ \ \ \ \ \ \ \ \ \ \ \ \ \ \ \ \ + \fr{1}{4 \skp^2 \fks} \l[ 5
k^2 F'^2 \l( 1 - \fr{27}{10} \sin^2 F \right) + 6 \sin^2 F ( k'^2 -
\omega^2 k^2 )  \right. \nn      \\
& & \ \ \ \ \ \ \ \ \ \ \ \ \ \ \ \ \ \ \ \ \ \ \ \ \ \ \ \ \  + 5 k^2
\sfrc \l(1 - \fr{7}{10} \sin^2 F \right)  + 9 k k' F' \sin 2F  \nn \\
& & \ \ \ \ \ \ \ \ \ \ \ \ \ \ \ \ \ \ \ \ \ \ \ \ \ \ \ \ \  + 3
\fr{k^2}{r^2}
\sin^2 F \cos^2\fr{F}{2} (1+ \fr{1}{3} \cos F )\Bigg] \ \Bigg\} \, ,  \\
\nn \\
\gamma_4^{(m)} &=& \fr{1}{30} \ e^2 \int dr \ r^4 \Bigg\{ k^2 \sin^2 F \nn \\
& & \ \ \ \ \ \ \ \ \ \ \ \ \ \ \ \ \ +
\fr{1}{4 \skp^2 \fks} \l[ \fr{9}{2} k^2
\sin^2 F \l( F'^2 + \fr{29}{9} \sfrc \right)  - 2 \sin^2 F ( k'^2 -
\omega^2 k^2 )  \right. \nn \\
& & \ \ \ \ \ \ \ \ \ \ \ \ \ \ \ \ \ \ \ \ \ \ \ \ \ \ \ \ \ - 3  k k' F'
\sin 2F  \nn \\
& & \ \ \ \ \ \ \ \ \ \ \ \ \ \ \ \ \ \ \ \ \ \ \ \ \ \ \ \ \  -
\fr{k^2}{r^2} \sin^2 F \cos^2\fr{F}{2} (1+ 27  \cos F )\Bigg] \ \Bigg\}
\, ,   \\
\nn \\
\gamma_5^{(m)} &=& \fr{1}{15} \ e^2
\int dr \ r^4 \Bigg\{ k^2 (2 - 3 \cos F) \nn \\
& & \ \ \ \ \ \ \ \ \ \ \ \ \ \ \ \ \ - \fr{1}{2 \skp^2 \fks} \l[ k^2 F'^2
\l( \fr{3}{2} \cos F - 1 \right) + \fr{k^2}{r^2} \sin^2 F \l( \cos F -
\fr{3}{2} \right)  \right. \nn \\
& & \ \ \ \ \ \ \ \ \ \ \ \ \ \ \ \ \ \ \ \ \ \ \ \ \ \ \ \ \ + 9 k k' F'
\sin F \Bigg] \ \Bigg\} \,  ,   \\
\nn \\
\gamma_6^{(m)} &=& -\fr{2}{15} \ e^2 \int dr \ r^4
\Bigg\{ k^2 \cos^2 \fr{F}{2} \nn \\
& & \ \ \ \ \ \ \ \ \ \ \ \ \ \ \ \ \ + \fr{1}{4 \skp^2 \fks} \l[ k^2 \cos^2
\fr{F}{2} \l( F'^2 + 14 \sfrc \right)  + 3 k k' F' \sin F \right] \ \Bigg\}\, .
\eea
$\gamma_1^{(e,m)}$ and $\gamma_2^{(e,m)}$  depend on the chiral angle
only,  while the remaining integrals
take into account the interplay between rotating soliton and bound kaon
wavefunction.

\section*{Appendix C}

For the sake of completeness we give in this Appendix the  explicit expressions
for the elementary magnetic moment operators  needed to calculate the
dispersive
contribution to the magnetic polarizability. The pure soliton contribution is
given by:
\bea
\mu_{s,0} & = &-{{2 M_N} \over {3 \pi \Theta}}
\int {\rm d}r \, r^2 \sin^2 F \, F' \, ,  \\
\mu_{v,0} & = & {1\over 2} M_N \Theta  \, .
\eea
The part describing the interplay between soliton field and bound kaon reads
\bea
\mu_{s,1} &=& c \ \mu_{s,0} - \frac{4}{3} M_N
 \int dr \  r^2 \left\{ k^2 \cos^2 \frac{F}{2} \right. \nonumber \\
& & \left. \hskip 0.5cm
+ \frac{1}{4 \skp^2 f_K^2} \left[ 4 \frac{k^2}{r^2} \sin^2 F
\cos^2 \frac{F}{2} + k^2 F'^2 \cos^2 \frac{F}{2} + 3 kk' F' \sin F \right]
\right\}, \label{19} \\
\mu_{v,1} &=& \frac{M_N}{3} \int dr \ r^2 \left\{ k^2 \cos^2
\frac{F}{2} \left( 1 - 4 \sin^2 \frac{F}{2} \right) \right. \nonumber \\
& & \left. \hskip 0.5cm + \frac{1}{4 \skp^2 f_K^2}
 \left[ 4 \frac{k^2}{r^2} \sin^2 F \cos^2 \frac{F}{2}
\left( 3 - 8 \sin^2 \frac{F}{2} \right) \right. \right. \nonumber \\
& & \left. \hskip 1.5cm + k^2 F'^2 \cos^2 \frac{F}{2}
\left( 1 - 18 \sin^2 \frac{F}{2} \right) - 2 k^2 \omega^2 \sin^2 F \right.
\nonumber \\
& & \left. \left. \hskip 1.5cm
 + 2 k'^2 \sin^2 F + 3 kk' F' \sin F \left( 3 - 4 \sin^2 \frac{F}{2} \right)
\right] \right\} \nonumber \\
& & + \frac{N_c M_N}{36} {\omega\over{f_K^2 \pi^2}}
\int dr \ r^2 \left( k^2 \sin^2 F F'
+ k k' \sin 2 F \right) \, .
\eea

\pagebreak

\begin{center}
\begin{tabular}{l|c|c|}
\cline{2-3}
  & Set I & Set II \\
\hline
$\!\!\! \vline \ \, \gamma^{(e)}_1$ &  20.7  &   32.1         \\
$\!\!\! \vline \ \, \gamma^{(e)}_2$ &  -10.4 &   -16.0        \\
$\!\!\! \vline \ \, \gamma^{(e)}_3$ &   0.78 &   1.11         \\
$\!\!\! \vline \ \, \gamma^{(e)}_4$ &   0.19 &   0.34         \\
$\!\!\! \vline \ \, \gamma^{(e)}_5$ &   2.14 &   4.11         \\
$\!\!\! \vline \ \, \gamma^{(e)}_6$ &  -2.23 &   -3.83        \\
\hline
\end{tabular}
\end{center}\begin{center}
{\large \bf Table I}: The elementary electric polarizabilities
(in $10^{-4} fm^3 $ ), as defined
in Appendix B, for Set I and Set II parameters.
\end{center}
\vspace{1.cm}

\begin{center}
\begin{tabular}{l|c|c|}
\cline{2-3}
  & Set I & Set II \\
\hline
$\!\!\! \vline \ \, \gamma^{(m)}_1$ &  -7.35 &  -11.1       \\
$\!\!\! \vline \ \, \gamma^{(m)}_2$ &  -3.00 &  -4.91       \\
$\!\!\! \vline \ \, \gamma^{(m)}_3$ &  -0.42 &  -0.62       \\
$\!\!\! \vline \ \, \gamma^{(m)}_4$ &   0.10 &   0.22       \\
$\!\!\! \vline \ \, \gamma^{(m)}_5$ &  -0.42 &  -0.93       \\
$\!\!\! \vline \ \, \gamma^{(m)}_6$ &  -0.72 &  -1.30       \\
\hline
\end{tabular}
\end{center}\begin{center}
{\large \bf Table II}: The elementary magnetic polarizabilities
in $10^{-4} fm^3 $ (seagull contribution).
\end{center}
\vspace{1.cm}

\begin{center}
\begin{tabular}{l|c|c|}
\cline{2-3}
  & Set I & Set II \\
\hline
$\!\!\! \vline \ \, \mu_{s,0}$ &  0.37  &  0.74          \\
$\!\!\! \vline \ \, \mu_{v,0}$ &  2.39  &  2.40          \\
$\!\!\! \vline \ \, \mu_{s,1}$ &  -1.11 &  -1.07         \\
$\!\!\! \vline \ \, \mu_{v,1}$ &  -0.10 &  -0.16        \\
\hline
\end{tabular}
\end{center}\begin{center}
{\large \bf Table III}: The elementary magnetic moments expressed in nuclear
magnetons (for more details,
see Ref.\cite{OMRS91}).
\end{center}

\pagebreak
\begin{center}
\begin{tabular}{l|c|c|}
\cline{2-3}
          & Set I & Set II \\
\hline
$\!\!\! \vline \ \, N$        &  17.3  &  26.7         \\
$\!\!\! \vline \ \, \Lambda$  &  18.1 &  28.0           \\
$\!\!\! \vline \ \, \Sigma^0$ &  18.1 &  28.0         \\
$\!\!\! \vline \ \, \Sigma^+$ &  18.8 &  29.4        \\
$\!\!\! \vline \ \, \Sigma^-$ &  17.4 &  26.5         \\
$\!\!\! \vline \ \, \Xi^0$    &  19.9 & 31.1          \\
$\!\!\! \vline \ \, \Xi^-$    &  18.0 &  27.3         \\
\hline
\end{tabular}
\end{center}\begin{center}
{\large \bf Table IV}: Electric polarizabilities (in $10^{-4} fm^3 $ ) for the
low lying octet
hyperons. Only the seagull contributions are here taken into account.
\end{center}
\vspace{1.cm}

\begin{center}
\begin{tabular}{l|c|c|c|c|c|c|}
\cline{2-7}
                &\multicolumn{3}{|c|}{Set I}
                &\multicolumn{3}{|c|}{Set II} \\
\cline{2-7}
           & $\beta_s$  & $\beta_d$ & $\beta_{tot}$
           & $\beta_s$  & $\beta_d$ & $\beta_{tot}$ \\ \hline
$\!\!\! \vline \ \, N$        &  -8.3 &  5.6 & -2.7
                              & -12.8 &  5.6  & -7.2  \\
$\!\!\! \vline \ \, \Lambda$  &  -8.7 & 12.1  & 3.4
                              & -13.3 &  12.0 & -1.3  \\
$\!\!\! \vline \ \, \Sigma^0$ &  -8.7 &  -4.0& -12.7
                              &  -13.3 &  -4.0 & -17.3  \\
$\!\!\! \vline \ \, \Sigma^+$ &  -9.1 &  10.4 & 1.3
                              & -14.0 &  10.1 & -3.9 \\
$\!\!\! \vline \ \, \Sigma^-$ &  -8.4 &   0.48 & -7.9
                              &  -12.6 &  0.12 & -12.5 \\
$\!\!\! \vline \ \, \Xi^0$    &  -9.6 &  14.0  & 4.4
                              &  -14.8 &  13.0 &  -1.8  \\
$\!\!\! \vline \ \, \Xi^-$    &  -8.7 &  1.5 & -7.2
                              & -13.0 &  0.59  & -12.4 \\  \hline
\end{tabular}
\end{center}\begin{center}
{\large \bf Table V}: Magnetic polarizabilities (in $10^{-4} fm^3 $ ) of octet
hyperons.
In this case, both seagull and dispersive parts contribute to the total
polarizability.
\end{center}

\end{document}